\documentclass[journal]{IEEEtran}
\usepackage{tikz}   
\ifCLASSINFOpdf
\else
\fi
\usepackage[cmex10]{amsmath}
\usepackage{amsthm}         

\newtheorem{theorem}{Theorem}[section]

\newtheorem{property}[theorem]{Property}
\newtheorem*{definition}{Definition}

\DeclareMathOperator{\bc}{bc}

\DeclareMathOperator{\im}{im}

\usepackage{enumerate}      
\usepackage{url}

\usepackage{graphicx}

\begin{document}
%
\title{Node Dominance: Revealing Community and Core-Periphery Structure in Social Networks}
%
%
%

\author{Jennifer~Gamble,~\IEEEmembership{Student~Member,~IEEE,}
        Harish~Chintakunta,~\IEEEmembership{Member,~IEEE,}
        Adam~Wilkerson,~\IEEEmembership{Member,~IEEE,}
        Hamid~Krim,~\IEEEmembership{Fellow,~IEEE}
        and Ananthram~Swami,~\IEEEmembership{Fellow,~IEEE,}
\thanks{Jennifer Gamble and Hamid Krim are with the Department
of Electrical and Computer Engineering, North Carolina State University, Raleigh, NC, USA, e-mail: jpgamble@ncsu.edu and ahk@ncsu.edu.}
\thanks{Harish Chintakunta is with the Coordinated Science Laboratory, University of Illinois Urbana Champaign, Champaign, IL, USA, email: hkchinta@illinois.edu.}
\thanks{Adam Wilkerson, Terrence J. Moore and Ananthram Swami are with the Army Research Lab, Adelphi, MD, USA. email: \textbf{add emails}}
}
\maketitle

\begin{abstract}
This study relates the local property of node dominance to local and global properties of a network. Iterative removal of dominated nodes yields a distributed algorithm for computing a core-periphery decomposition of a social network, where nodes in the network core are seen to be essential in terms of network flow and global structure. Additionally, the connected components in the periphery give information about the community structure of the network, aiding in community detection. A number of explicit results are derived, relating the core and periphery to network flow, community structure and global network structure, which are corroborated by observational results. The method is illustrated using a real world network (DBLP co-authorship network), with ground-truth communities.
\end{abstract}


%
\IEEEpeerreviewmaketitle

\section{Introduction}
%
%
%
%


\IEEEPARstart{O}{NE} of the interesting challenges in social networks is to relate local connectivity properties to global structure. The motivation for doing do stems from the belief that local  properties reflect interactions amongst individuals (or entities). Therefore such relationships help us make inferences about the nature of interactions which led to the network, by studying its global properties.      {I}{n} this paper, we present the local property of \emph{node dominance} as a method for network analysis. We will show why node dominance is such a useful criterion, by developing a low complexity, distributed algorithm for the core-periphery decomposition of a network based on node dominance criteria. We will also demonstrate its relation to the network community structure.

Owing to a localized definition, the node dominance criteria for a node v can be determined only from a two hop neighborhood. A node $v$ is dominated by node $w$ if all nodes that share and edge with $v$, also share an edge with $w$. The formal definition of node dominance is based on a simplicial complex (as opposed to graph) structure, and will be discussed in detail later. If we iteratively collapse dominated nodes, the resulting set (the network \emph{core}) is shown to consist of nodes that are important with respect to the network flow, community structure, and global network structure. One especially important property of the core is the preservation of shortest distances, so a shortest path between any two nodes in the core is also a shortest path between them in the original network. The network \emph{periphery} (the complement to the core, consisting of dominated nodes) is seen to consist of many connected components, including all the nodes in the network through which no shortest paths pass. These peripheral components also play a key role in the community structure of the network.

The intuitive notion that a network naturally decomposes into a core and periphery has appeared many times in the social network literature over the decades. Researchers have proposed different interpretations about what such a decomposition should look like, but it is commonly suggested that a `core' should be central to the network (with respect to information flow, or shortest paths) \cite{Holme2005}, have high average degree \cite{Csermely2013}, and be relatively well-connected both internally, and to the periphery \cite{Borgatti2000} \cite{Zhang2014}. In contrast, the periphery should be connected to the core, but extremely sparsely connected amongst itself.

Borgatti and Everett \cite{Borgatti2000} were the first to attempt to analytically describe these intuitive properties. They proposed an `idealized core-periphery', wherein every core node is connected to every other core node, each peripheral node is connected to the core, and no peripheral nodes are connected to each other. They would then learn the core-periphery structure for a given network by assigning each node as `core' or `periphery' in the way that best correlated with this idealized structure. This method assumes explicitly that the probability of two nodes being joined by an edge is only a function of their `core-ness', as opposed to some other characteristics, such as community membership. In this sense, the core-periphery model considered in \cite{Borgatti2000} is in contrast to  common network models based on community structure. Both core-periphery and community network structures can be expressed using a stochastic blockmodel approach \cite{Zhang2014}, but with different parameters, so under these models a given network will not display both structures simultaneously.

Another approach, by Rombach \emph{et al.} \cite{Rombach2014} presents a  generalization of Borgatti and Everett's philosophy, where a \emph{core score} is computed for each node, using a range of possible core sizes and continuous/discrete transitions between core and periphery. Here, they admit that both core-periphery and community structure are often present in real-world networks, but still propose the core-periphery decomposition as an alternative/complementary analysis to the more common community detection methods. In Della Rossa \emph{et al.} \cite{DellaRossa2013}, an approach to periphery detection based on random walks is taken, where is it assumed that due to the extremely sparse connectivity of the periphery, a random walk will exit the set of peripheral nodes very quickly. Thus, a core-periphery profile for the network, along with a coreness value for each node, is computed using a greedy algorithm that incrementally adds nodes to the periphery in a way that minimizes the expected exit time of a random walk. Again, this method focuses very heavily on the sparsity of the periphery, and is somewhat unrelated to any community structure that may be present in the network. For a good review of existing methods of core-periphery network decomposition, see the survey by Csermely \emph{et al.} \cite{Csermely2013}, or the introductory sections in \cite{Rombach2014}.

Traditionally, approaches to community detection in networks have assumed that communities form a partition of the network, with each node belonging to exactly one community. A foundational method has been the Girvan-Newman algorithm \cite{Newman2004}, where communities are detected though iterative removal of edges with high betweenness centrality. They defined the notion of `modularity' as a stopping criterion for their algorithm, and many subsequent algorithms attempt to partition a network in such a way that optimizes (usually approximately) modularity \cite{Newman2004b}, or cut ratio (approximated using spectral clustering) \cite{Chan1994}.  Fortunato provides an excellent overview of the breadth and depth of approaches to the community detection problem in his 100 page survey paper \cite{Fortunato2010}. In more recent years, researchers are determining that partition-based methods are often somewhat unrealistic, since real-world networks with ground-truth communities typically display overlapping community structure \cite{Yang2012}, where one node may have multiple community memberships. See Xie \emph{et al.} \cite{Xie2013} for a survey of methods for overlapping community detection, including clique percolation, link clustering, and fuzzy detection methods using mixed-membership stochastic block models, or nonnegative matrix factorization.

A particularly realistic model for overlapping community detection is Yang and Leskovec's community-affiliation graph model (AGM) \cite{Yang2012b} \cite{Yang2014}. This model considers communities as `overlapping tiles', and its distinguishing feature is that regions of community overlaps are \emph{more} densely connected than regions involving single communities. Precisely, the probability of an edge existing between two vertices is based on the communities they share, with higher probability when they have more community memberships in common. This assumption is validated on data sets with ground-truth community memberships available, where higher edge densities are observed in community intersections \cite{Yang2012b}. AGM, and the other methods for overlapping community detection are more realistic than the partition-based methods, but they do not scale up well with size of the network. A recent relaxation of AGM, referred to as Cluster Affiliation Model for Big Networks (BIGCLAM) \cite{Yang2013}, allows nodes to have continuous-valued community memberships, indicating their degree of involvement in a given community. This reduces the combinatorial optimization in AGM to a continuous optimization that can be solved using nonnegative matrix factorization, making it viable for large networks. We will return to these models in Section \ref{CommDetec}.

In the current paper, we will see how a core-periphery structure and a community structure are both present in real-world networks, and how node dominance informs us about both. The relationship between the core-periphery and community structure of a network has been touched upon previously by Leskovec \emph{et al.} \cite{Leskovec2009}, where they also noted the presence of a network periphery, defined in terms of \emph{whiskers} (clusters of nodes that are separable from the main network by removing a single edge), which were interpreted as small communities, weakly connected to the remaining network ``core''. In the AGM model mentioned above \cite{Yang2014}, Yang and Leskovec refer to the overlapping portions of communities as the ``core'' of the network. We will see that this interpretation does in fact concur with our notion of core and periphery, where in networks with ground-truth communities available, the nodes in the core obtained using node dominance typically have multiple community memberships, while the nodes in the periphery have fewer community memberships (often just one).

Iterative node dominance collapses were originally proposed independently by Wilkerson \emph{et al.} \cite{Wilkerson2013} and Barmak and Minian \cite{Barmak2012}, as a homology/homotopy-preserving simplification of a simplicial complex, with the distributed version described in \cite{Wilkerson2013b}. Here, we explore much more deeply the use of this simplification as a network core, and describe the relationship between the core-periphery decomposition, and the community structure, global structure, and network flow properties.

In Section \ref{Background}, we will first describe the relevant information for the simplicial complex representation of a network, and the background and definition of the node dominance criterion. We follow this in Section \ref{Properties} by statements and derivations of the resulting properties of  core-periphery decomposition, and present an algorithm for the use of peripheral components in community detection. In Section \ref{Exp}, we illustrate our method with two real-world network data sets which contain ground-truth community information. We not only empirically verify the importance of core nodes with respect to network flow and global structure, but see that our propose d use of the peripheral components for community detection outperforms BIGCLAM, which is considered the current state-of-the-art method for overlapping community detection in large networks. Finally, in Section \ref{CorePeriphConcl} we draw some conclusions, and discuss the limitations of our method, as well as some directions for future research.


\section{Background}\label{Background}

\subsection{Simplicial homology}

A graph $G = G(V,E)$ is defined by a list, $V$, of its vertices, as well as a list, $E$, of the pairs of vertices that are joined by an edge. An implicit assumption in this is that an edge $e = (v_i,v_j) \in E$ can only be present in $G$ if both of its vertices $v_i$ and $v_j$ are in $V$. The notion of a simplicial complex is a higher-order generalization of a graph, while similarly preserving this `closed under subsets' property.

\begin{definition}[Simplicial complex]
A \emph{$k$-simplex} $\sigma = (v_0,v_1,\ldots,v_k)$ is a set of $(k+1)$ singleton elements (called \emph{vertices}). A \emph{simplicial complex} $K$ is a set of simplices (i.e. a set of sets of vertices) such that
\begin{enumerate}[(i)]
\item if $\sigma, \tau \in K$, then $\sigma \cap \tau \in K$
\item if $\tau \leq \sigma$, then $\tau \in K$
\end{enumerate}
where $\leq$ indicates the subset relation. If $\tau \leq \sigma$, we call $\tau$ a \emph{face} of $\sigma$.
\end{definition}

A simplex $\sigma$ is \emph{maximal} if there are no $\tau \in K$ such that $\sigma < \tau$. A $k$-simplex has dimension $k$. The \emph{dimension} of simplicial complex $K$ is the maximum dimension of any simplex in $K$
\[ \dim(K) = \max_{\sigma \in K} \dim(\sigma). \]
A subset $K'$ of a simplicial complex $K$ is called a \emph{subcomplex}, if $K'$ is itself a simplicial complex (satisfying properties (i) and (ii) above). The $k$-skeleton of $K$ is the subcomplex formed by all simplices in $K$ with dimension at most $k$
\[ k\mbox{-skeleton of } K = \{\sigma \in K \mbox{ } | \mbox{ } \dim(\sigma) \leq k \} \]

\begin{definition}
Let $K_1$ and $K_2$ be two simplicial complexes with vertex sets $V_1$ and $V_2$. A map $\phi_0: V_1 \rightarrow V_2$ on the vertex sets induces a \emph{simplicial map} $\phi:K_1\rightarrow K_2$ on the complexes, if for every simplex $\sigma = (v_0,\ldots,v_k)\in K_1$, the set $(\phi_0(v_0),\ldots,\phi_0{v_k})$ spans a simplex in $K_2$. A simplicial map $\phi:K_1 \rightarrow K_2$ induced by an isomorphic map on the vertex sets is said to be an \emph{isomorphic simplicial map}, and in this case, $K_1$ and $K_2$ are \emph{isomorphic simplicial complexes}.
\end{definition}

In Section \ref{Properties}, this isomorphism between complexes will be used to describe the uniqueness of the core obtained using node dominance collapsing. 

Given a graph $G = G(V,E)$, we can think of $G$ as the 1-skeleton of a simplicial complex, whose higher-dimensional simplices have not been directly observed. The maximal simplicial complex whose 1-skeleton is equal to $G$ is called the \emph{flag complex}.

\begin{definition}[Flag complex]\label{flag}
Given a graph $G = G(V,E)$, the simplicial complex
\begin{align*}
X(G) = \{ \sigma & = (v_{i_0},v_{i_1},\ldots,v_{i_{\dim\sigma}})  \mbox{ } | \mbox{ } \\
                 & (v_{i_j},v_{i_k}) \in E \mbox{ for all } 0\leq j,k \leq \dim\sigma \}
\end{align*}
contains a simplex $\sigma$ whenever all pairs of vertices in $\sigma$ are connected by an edge in $E$. $X(G)$ is called the \emph{flag complex} of $G$.
\end{definition}

As we will see in Section \ref{HomRel}, if we have additional information about the $k$-tuple relations in $G$, we may build a simplicial complex using that information, adding $k$-simplex $\sigma$ whenever its vertices satisfy a $k$-tuple relation, and all faces of the simplex are also present. In the absence of such information, when only the graph $G$ is given, we propose the use of the flag complex, and see that it can be very informative.

A final notion we will mention here is the definition of the \emph{homology} of a simplicial complex.

\begin{definition}[Homology]\label{Homology}
We encode the structure of simplicial complex $X$ through \emph{boundary maps} $\{\partial_k\}_{k=1}^{\dim(X)}$, where $\partial_k$ gives the oriented connectivity information between $k$-simplices and $(k-1)$-simplices. Then the $k$-th homology group of $X$ is
\[ H_k(X) = \ker(\partial_k)/\mbox{ }\im(\partial_{k+1}) \]
See, for example, \cite{Hatcher2002} for a more mathematically complete definition of simplicial homology.
\end{definition}

Intuitively, the dimension of the $k$-th homology space counts the number of $k$-dimensional ``holes'' in the simplicial complex. These can be thought of as $(k+1)$-dimensional voids enclosed by $k$-simplices, so $H_1$ counts the number of loops which are not ``filled-in'' by triangles, and $H_2$ counts the number of voids. The interpretation of $H_0$ is slightly different: it counts the number of connected components of $X$ (which may be interpreted as cycles of dimension zero).

The sequence of homology spaces of a simplicial complex, in essence, specify the "global structure" of the complex. For our purposes, we will not be computing any homology directly, but we will see that by preserving homology during our node dominance collapse, we will in fact be preserving important global structure of the network.

\subsection{Node dominance}\label{dominance}

We will be representing a network using its flag complex, and in that setting, \emph{node dominance} is characterized by the following definition.

\begin{definition}
The \emph{neighbor set} of a node $v$, is the set of all nodes sharing an edge with $v$, as well as $v$ itself:
\[ \mathcal{N}[v] := \{u \in V \mbox{ } | \mbox{ } (u,v) \in E\} \cup \{v\}. \]
A node $v$ is \emph{dominated by} one of its neighbors $w$, if and only if $\mathcal{N}[v] \subseteq \mathcal{N}[w]$ i.e., all the neighbors of $v$ are also neighbors of $w$.
\end{definition}

To understand the importance and relevance of this definition, we will explore a bit of its history, and related concepts.

\subsubsection{Homology of a relation}\label{HomRel}

\begin{definition}
A \emph{relation} on two sets $A$ and $B$ is a function $r: A\times B \to \{0,1\}$. We say that elements $a_i, a_j \in A$ are \emph{related} (through element $b$) if there exists an element $b \in B$ such that $r(a_i,b) = 1$ and $r(a_j,b) = 1$. Similarly, $b_i, b_j \in B$ are related if there exists an $a \in A$ such that $r(a,b_i) = 1$ and $r(a,b_j) = 1$. For $A$ and $B$ finite, the relation $r$ can be represented by an $|A|\times|B|$ binary matrix $R =(r_{ij})$, where $r_{ij} = r(a_i,b_j)$.
\end{definition}

As an example, the elements of set $A$ could be actors, and the elements of set $B$ could be movies, with $r(a,b) = 1$ whenever actor $a$ appears in movie $b$.

Given a relation, there are two ways to encode its structure as a simplicial complex. The first way, which we will denote as $X_R(A,B)$, the elements of $A$ are represented as vertices, and vertices $\{a_{i_0}, a_{i_1}, \ldots, a_{i_k}\}$ are spanned by a $k$-simplex whenever there exists a $b \in B$ such that $r(a_{i_l},b) = 1$ for all $l = 0, 1, \ldots, k$. The second way, which we will denote as $X_R(B,A)$, the elements of $B$ are represented as vertices, and $\{b_{j_0}, b_{j_1}, \ldots, b_{j_k}\}$ are similarly spanned by a $k$-simplex whenever they are all related by the same $a \in A$. Note also that for any simplicial complex $X$ (even if it wasn't constructed using a relation) one may form its dual complex $\hat{X}$, by letting each maximal simplex in $X$ correspond to a vertex in $\hat{X}$. In that case, a set of vertices in $\hat{X}$ are spanned by a simplex if their associated simplices in $X$ all had a vertex in common.

In the example with actors and movies, this means that we can represent their relationships by building a simplicial complex where actors are vertices, and simplices are formed between actors who are in the same movie; or alternatively, we can encode it by using movies as vertices and spanning a set of movies by a simplex when they all feature the same actor.

Note that these two simplicial complexes may have drastically different structure (different number of vertices, different dimension), but Dowker \cite{Dowker1952} proved that the two complexes have exactly the same homology (in the sense that the $k^{th}$ homology groups of the two complexes are isomorphic, for all $k$).

\begin{theorem}[Dowker]
If $R$ is a relation on sets $A$ and $B$, with associated simplicial complexes $X_R(A,B)$ and $X_R(B,A)$, then
\[ H_k(X_R(A,B)) \cong H_k(X_R(B,A)) \mbox{ for all } k \]
\end{theorem}

\subsubsection{Node dominance and equivalent notions}\label{Equiv}

In light of the dual simplicial complexes presented in Section \ref{HomRel}, we can now give the more general definition of node dominance.

\begin{definition}[Node dominance]
Given simplicial complex $X$ and its dual complex $\hat{X}$, each vertex $v \in X$ has an associated simplex $\sigma_v \in \hat{X}$. We say a vertex $v$ is \emph{dominated} by vertex $w$, if $\sigma_v$ is a face of $\sigma_w$. This occurs exactly when the set of simplices incident to (i.e. containing) $v$ is a subset of the set of simplices incident to $w$ (in $X$).
\end{definition}

When the simplicial complex of interest is a flag complex, we know that the presence of a higher dimensional simplex is determined by the presence of its constituent edges. This is why we are able to check the node dominance criterion using only the neighbor sets of our vertices, in the flag complex setting: if the neighbors of $v$ are all neighbors of $w$, then the set of simplices incident to $v$ is a subset of the set of simplices incident to $w$.

To illustrate the concept of node dominance using the example of actors and movies, consider two actors, represented by separate vertices $a_i$ and $a_j$ in $X_R(A,B)$. If the movies featuring actor $a_i$ is a (proper) subset of the movies featuring actor $a_j$ (i.e. $a_i$ is dominated by $a_j$), then in the dual complex $X_R(B,A)$, the simplex $\sigma_{a_i}$ will be a (proper) face of simplex $\sigma_{a_j}$. Thus, removing actor $a_i$ (and all its incident simplices) completely, will not change the simplicial structure of the dual complex $X_R(B,A)$ at all, and thus will not change the homology of the original complex $X_R(A,B)$.

The insight that removing dominated nodes does not change the homology of the simplicial complex, suggests an algorithm, as originally proposed (independently) by \cite{Wilkerson2013} and \cite{Barmak2012}, to simplify a simplicial complex by iteratively removing such vertices. In the work by Barmak and Minian \cite{Barmak2012}, they term the removal of a dominated node a \emph{strong homotopy collapse}, node dominance is a stricter condition than that required for a regular homotopy-preserving simplicial collapse \cite{Whitehead1939}.

\begin{figure}[htp]
\begin{center}
\begin{tabular}{ll}
\includegraphics[scale=0.15]{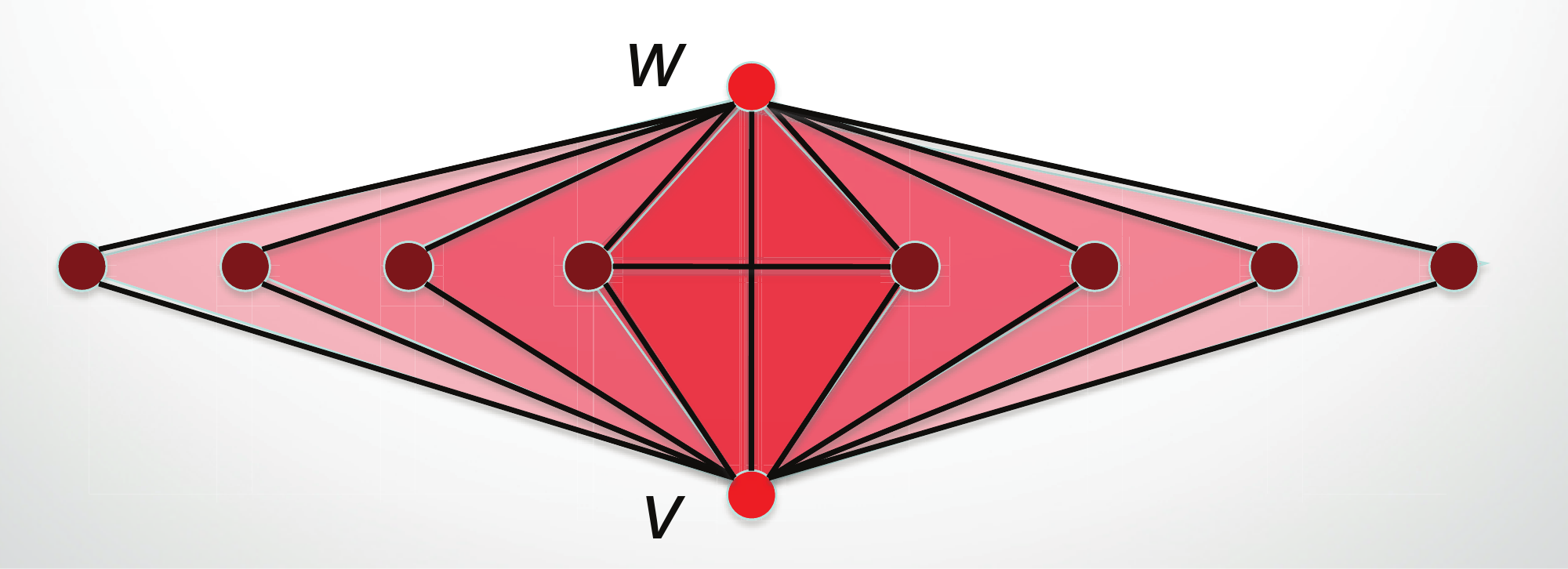} & \includegraphics[scale=0.15]{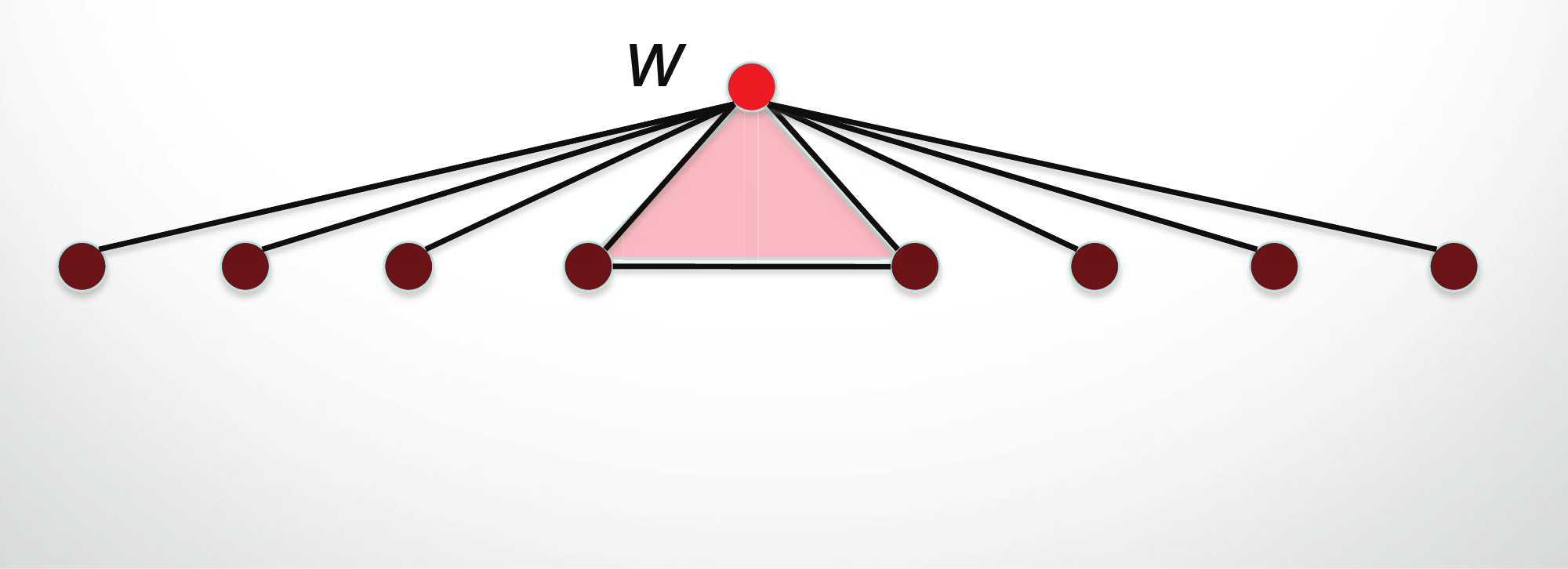} \\
\end{tabular}
\end{center}
\caption{Node $v$, dominated by node $w$. Removal of $v$ only has local effects.\label{Dominated}}
\end{figure}

In Figure \ref{Dominated}, vertex $v$ is dominated by vertex $w$, where vertex $w$ could have additional connections in the network which are not shown. The removal of vertex $v$ does not create or destroy any connected components, loops, or voids (preserves homology), and does not affect shortest path lengths between other nodes (see Section \ref{Flow}).

One more definition we will note is that of a \emph{2-hop neighbor set}, which is the neighbor set of a node that also contains all ``friends of friends'', instead of just immediate neighbors:
\small{
\[ \mathcal{N}_2[v] = \{ u \in V \mbox{ } | \mbox{ } (u,v) \in E, \mbox{ or } (u,v_i) \in E \mbox{ for some } v_i \in \mathcal{N}[v] \} \]
}
Performing the node dominance collapse using the 2-hop neighbor set can allow greater collapsability in networks with few dominated nodes. It also allows small holes in the flag complex (i.e. those with hop length $\leq$ 6) to be ``filled in'', so only larger homological features are preserved. We will use this version of the node dominance collapse on one of the data sets in Section \ref{Exp}.

\subsubsection{Distributed algorithm for flag complexes}\label{Algorithm}

Assuming a flag complex structure, the node dominance collapse can be performed referring only to its 1-skeleton (the original graph under analysis). Moveover, the criterion for determining node dominance requires only local information, making the algorithm of distributed nature. This algorithm was first presented in \cite{Wilkerson2013b}.

Each node $v$ has the list of its neighbor set $\mathcal{N}[v]$, and it then executes the following steps during each iteration:

\vspace{5mm}
\boxed{
\begin{array}{l}
\textbf{Distributed algorithm for node dominance collapse} \\
\mbox{Broadcast } \mathcal{N}[v] \mbox{ to neighbors} \\
\textbf{for } v_i \in \mathcal{N}[v], v_i \neq v\\
\indent \mbox{Receive } \mathcal{N}[v_i] \\
\indent \textbf{if } \mathcal{N}[v_i] \subseteq \mathcal{N}[v] \\
\indent \indent \mbox{Broadcast OFF to } v_i \\
\indent \indent \textbf{if } \mbox{OFF received from } v_i \\
\indent \indent \indent \mbox{Handshake to determine if } v \mbox{ or } v_i \mbox{ turns off} \\
\indent \indent \textbf{end if} \\
\indent \textbf{end if} \\
\textbf{end for} \\
\textbf{if } \mbox{OFF received OR Handshake determined } v \mbox{ turns off} \\
\indent v \mbox{ designated OFF} \\
\textbf{else} \\
\indent \mbox{Update } \mathcal{N}[v] \mbox{, omitting OFF neighbors} \\
\end{array}
}
\vspace{5mm}

A very similar distributed algorithm is also possible in the non-flag complex setting, where there exists some \emph{a priori} information about which $k$-tuples of simplices are related. An example of this would be the list of movies and actors, or some other relation (eg. authors/papers). In that case three actors (vertices) are only spanned by a triangle when there is a single movie they all appeared in together, not only if they had all appeared in movies together pairwise, as in the flag complex case. To compute node dominance in that setting, we only need to assume that each node has access to its \emph{list of maximal simplices} (eg. an actor has its movie list, an author has its paper list, etc.). Then the algorithm above can proceed exactly as written, with $\mathcal{N}[v]$ replaced by the maximal simplex list of $v$.

\section{Properties of core and periphery}\label{Properties}

In this section, we will outline both the analytical and empirically observed properties of the core-periphery decomposition obtained through the iterative node dominance collapse. Examples of the observed properties on real-world data sets are presented in Section \ref{ObsProp}.

\textbf{Analytical properties:}
\begin{enumerate}
\item Shortest paths in the core are shortest paths in the original network. \textit{(Network flow)}
\item Nodes with betweenness centrality zero are not in the core \textit{(Network flow)}
\item A node is more likely to be dominated by a node sharing the community membership(s) of its neighborhood set, compared to a node which does not. \textit{(Community structure)}
\item The homology of the flag complex of the core is the same as the homology of the flag complex of the entire network \textit{(Global structure)}
\item The structure of the core is unique (all possible cores for a given network are isomorphic as simplicial complexes) \textit{(Global structure)}
\end{enumerate}

\textbf{Observed properties:}
\begin{itemize}
\item Core nodes typically have high degree and high betweenness centrality. `Hub' nodes are in the core. \textit{(Network flow)}
\item Nodes with multiple ground-truth community membership labels tend to be in the core, while nodes with just one (or no) community labels are usually in the periphery. \textit{(Community structure)}
\item Using the peripheral groups, we can obtain candidate sets that are seen to contain a large proportion of ground-truth communities. See Section \ref{CommDetec} for details, and our use of these candidate sets for community detection. \textit{(Community structure)}
\item The core is stable with respect to the order of collapses in the iterative algorithm.  \textit{(Global structure)}
\end{itemize}

Throughout this section, for a graph $G = G(V,E)$, the core $G_C = G(V_C,E_C)$ is the graph induced by the set of nodes $V_C \subseteq V$ which remain upon an iterative and total removal of dominated nodes from $V$. Note that the set $V_C$ (and thus the core itself) is not necessarily unique, because of a potential random `handshake' in the Algorithm from Section \ref{Algorithm}. The statements given below are valid for any core obtained by the procedure of iterative node dominance collapse. As we will discuss further in Section \ref{Global} below, all possible cores obtained from the same initial graph have the exact same structure (are isomorphic) \cite{Matouvsek2008}.

\subsection{Network flow}\label{Flow}

The properties in this subsection involve statements about shortest paths between given nodes in the network. An outline of a proof similar to Property \ref{ShortestPaths} is given in \cite{Wilkerson2013}, and we include the  proof here for completeness.

\begin{definition}[Shortest paths]
Given a graph $G' = G(V,E)$, for any pair of points $v_i,v_j,\in V$, a \emph{path} $p = (v_i = v_1, v_2, \ldots, v_l = v_j)\footnote[1]{Note that there is no loss of generality by using indices 1,2,\ldots,l}$ is a sequence of vertices such that $(v_k,v_{k+1}) \in E$ for all $k = 1,\ldots, l-1$. The path has length $|p| = l$, and $p$ is a \emph{shortest path} if $l \leq |p'|$ for any other path $p'$ from $v_i$ to $v_j$. The set of all shortest paths from $v_i$ to $v_j$, in the graph $G'$ is denoted $SP_{G'}(v_i,v_j)$.
\end{definition}

\begin{property}[Shortest paths in the core are shortest paths in the original network.]\label{ShortestPaths}
For $v_1, v_2 \in V_C$, if $p \in SP_{G_C}(v_1,v_2)$, then $p \in SP_G(v_1,v_2)$.
\end{property}
\begin{proof}
For any graph $G'$, let $v_j$ be dominated by its neighbor $v_i$. Consider any shortest path $p = (\ldots,v_k,v_j,v_l,\ldots)$ passing through $v_j$. Note that $k,l\neq i$ [Proof by contradiction: $p = (\ldots,v_i,v_j,v_l,\ldots)$ could be replaced by shorter path $(\ldots,v_i,v_l,\ldots)$, since $\mathcal{N}[v_j] \subseteq \mathcal{N}[v_i]$ so $v_l \in \mathcal{N}[v_j] \Rightarrow v_l \in \mathcal{N}[v_i]$]. So $p = (\ldots,v_k,v_j,v_l,\ldots)$ can be replaced by $p' = (\ldots,v_k,v_i,v_l,\ldots)$, which is the same length as $p$, but doesn't contain $v_j$. \\
Therefore, the length of all shortest paths in $G'$ (where $v_j$ is not the source or destination) are preserved when $v_j$ is removed.
\end{proof}

\begin{definition}[Betweenness centrality]
The betweenness centrality of a node $v$  is defined as the proportion of shortest paths between nodes $s$ and $t$ that pass through $v$, summed over all pairs $s,t\neq v$. i.e.)
\[ \bc(v) = \sum_{s,t\neq v} \frac{|\{p \in SP_G(s,t) | v \in p\}|}{|SP_G(s,t)|} \]
\end{definition}

\begin{property}[If the size of the core is greater than 1\footnote{In practice, this assumption is almost always satisfied.}, nodes with betweenness centrality zero are not in the core]\label{bcProp}
\[ \bc(v) = 0 \Rightarrow v \not\in V_c \]
\end{property}
\begin{proof}
Using the definition of betweenness centrality above, we can see that
\[ bc(v) = 0 \Rightarrow |\{p \in SP_G(s,t) | v \in p\}| = 0 \mbox{ }\forall s,t\neq v. \]
Therefore, either
\begin{enumerate}[(i)]
\item $\deg(v) = 1$
\item $\forall s,t, \in \mathcal{N}[v], (s,t) \in E$ (so that $\ldots,s,v,t,\ldots$ will not be in any shortest path)
\end{enumerate}
If (i), then $v$ is dominated. \\
If (ii), then $\mathcal{N}[v]$ is a clique, so for any $w \in \mathcal{N}[v]$ with $w \neq v$, $\mathcal{N}[v] \subseteq \mathcal{N}[w]$. This implies $v$ is dominated by all its neighbors. In this case, either $v$ is removed and therefore in the periphery, or all its neighbors are removed and $v$ is the only node in the core. Since we assume that the size of the core is greater than 1, $v \not\in V_C$.
\end{proof}

Both of these properties speak to the `centrality' of the nodes in the core, with respect to the original network. Property \ref{ShortestPaths} tells us that there is no way to shortcut through the periphery when traveling between two nodes in the core, and Property \ref{bcProp} says the nodes that are not involved in any shortest paths are guaranteed to be contained in the periphery. Together, we can conclude that the node dominance collapse only has local effects (with respect to shortest paths in the network), in that only shortest paths beginning or ending at the dominated node are affected.

Empirically, we see that nodes with high betweeness centrality and nodes with high degree will lie in the core (see Section \ref{ObsProp} for concrete examples). These are `hub' nodes, in terms of network flow properties, so removal of nodes in the core have a much greater impact on network information flow than removal of nodes from the periphery.

\subsection{Community structure}\label{CommStruc}

The community affiliation graph model (AGM) proposed by Yang and Leskovec \cite{Yang2012b} assumes that the probability of an edge forming between two nodes depends on the community membership(s) of the nodes under consideration. This is similar to the traditional stochastic blockmodel (which require communities to form a partition of the network), or generalizations \cite{Airoldi2009} of the stochastic blockmodel that allow for overlapping communities, with the notable exception that under AGM the edge density in the intersections of communities is \emph{higher} than the edge density in the non-overlapping portions of communities.

For notation, consider the set $C = \{c_k\}_{k=1}^m$ defining the $m$ communities in the network, where $c_k$ is the set of nodes belonging to the $k^{th}$ community. Note that each node in $V$ may belong to zero, one, or multiple communities. For two nodes $u,v \in V$, let $C_{uv} = \{c \in C \mbox{ }| \mbox{ }u,v \in c\}$ denote the set of communities containing both $u$ and $v$. We will also use the more general notation $C_S = \{ c \in C \mbox{ } | \mbox{ } \exists v \in S \mbox{ s.t. } v \in c\}$ to denote the set of community memberships for nodes in a given set $S$. Under AGM, an edge forms between $u$ and $v$, independently, with probability $p_c$ for each of the communities $c \in C_{uv}$. In other words, denoting the probability of an edge between $u$ and $v$ by $p(u,v) = P[(u,v) \in E]$, we have
\begin{equation}\label{AGMedge}
p(u,v) = 1 - \prod_{c \in C_{uv}} (1 - p_c).
\end{equation}
Further, Yang and Leskovec define a baseline edge probability $\varepsilon = p(u,v)$ for $u,v$ with no communities in common. They choose $\varepsilon = \frac{2|E|}{|V|(|V|-1)}$, which is typically a number of orders of magnitude smaller than the $p_c$ probabilities. For the proof of the following result, we assume the AGM model for network community structure, however the result would still hold for any model that bases the probability of an edge between two nodes on the community membership of the nodes, where the probability of an edge is significantly higher for nodes sharing communities than nodes not sharing communities.

\begin{property}[A node is more likely to be dominated by a node sharing the community membership(s) of its neighborhood set, compared to a node which does not.]\label{commProp}
In other words, $v$ is dominated by $w$ with much higher probability when $C_{\mathcal{N}[v]} \subseteq C_w$ as compared to the case when $C_{\mathcal{N}[v]} \not\subseteq C_w$
\end{property}
\begin{proof}
The probability that $v$ is dominated by $w$ is
\footnotesize
\begin{eqnarray*}
P[v \mbox{ dom. by } w] &{=}& \prod_{v_i \in \mathcal{N}[v]} p(w,v_i) \\
&{=}& \left(\prod_{\substack{v_i \in \mathcal{N}[v]\\ C_{wv_i} \neq \emptyset}}\left[1 - \prod_{c \in C_{wv_i}}(1 - p_c)\right] \right) \prod_{\substack{v_i \in \mathcal{N}[v]\\ C_{wv_i} = \emptyset}} \varepsilon \\
\end{eqnarray*}
\normalsize

In other words, $v$ will be dominated by $w$, only if there exist edges between $w$ and all $v_i \in \mathcal{N}[v]$. Each of these edges occurs independently, with probability $p(w,v_i)$, with the value given in Equation (\ref{AGMedge}) if $w$ and $v_i$ share community membership(s) (i.e. if $C_{wv_i} \neq \emptyset$), and $p(w,v_i) = \varepsilon$ otherwise. \\
Since $\varepsilon \ll p_k$ for all $k$,
\[ P[(w,v_i) \in E \mbox{ } | \mbox{ } C_{wv_i} \neq \emptyset] \gg  P[(w,v_i) \in E \mbox{ } | \mbox{ } C_{wv_i} = \emptyset] \]
Therefore
\begin{align*} 
P[v \mbox{ dom. by } w & \mbox{ } | \mbox{ } C_{\mathcal{N}[v]} \subseteq C_w] \gg \\
                       & P[v \mbox{ dominated by } w \mbox{ } | \mbox{ } C_{\mathcal{N}[v]} \not\subseteq C_w] 
\end{align*}
\end{proof}

In real world networks (as described in Section \ref{ObsProp}), nodes in the periphery typically have one (or no) community membership(s), while nodes in the core have multiple community memberships, and lie in the intersections of communities. In Section \ref{CommDetec}, we will take this interpretation further, by proposing a method for using the peripheral components to obtain candidate sets which are likely to contain communities of the network. We can think of the peripheral components as the non-overlapping portions of the communities, in which case the true network communities would consist of a peripheral component, along with adjoining nodes in the core. It is also possible that a single community could have non-overlapping portions which ``stick out'' from the core in multiple places, on account of which we propose a method of combining peripheral components according to which core nodes they connect to. This yields an algorithm for obtaining ``candidate sets'' which are intended to contain the true network communities. This method is discussed further in Section \ref{CommDetec}.

\subsection{Global structure}\label{Global}

As described in Section \ref{dominance}, when the flag complex representations of the original network and the core network are used, the core is seen to have the exact same homology as the original complex, in the sense that their homology spaces are isomorphic in all dimensions.

\begin{property}[Homology is preserved in the core]\label{homProp}
\[ H_k(X(G_C)) \cong H_k(X(G)) \mbox{ for all } k \]
\end{property}
\begin{proof}
This property follows immediately from the result of Dowker's Theorem (that a simplicial complex and its dual complex have the same homology), combined with the observation that if a vertex is dominated, its corresponding simplex in the dual complex will be a face of the simplex corresponding to the dominating node, and thus will not contribute to the structure of the dual complex. \\
An alternative formulation and proof is available in \cite{Barmak2012}.
\end{proof}

A corollary of Property \ref{ShortestPaths} is that at least one shortest cycle for each homology class is retained in the core. Thus, not only is the dimension of each homology space preserved, but the `hole locations' in the network are also preserved. It is this additional property that truly allows us to interpret the core as the global scaffolding for the network.

Property \ref{homProp}, together with Property \ref{commProp} tell us that nodes with diverse friend sets (including bridging ties) will be in the core. If they are not, it is only because they are dominated by another node with all the same diverse connections. In real-world networks, we see that the average clustering coefficient for nodes in the core is much lower than in the network as a whole (see Section \ref{ObsProp}), which supports the `diverse friend set' interpretation, because the friends of a core node are usually not friends with each other.

\section{Analysis of real-world networks}\label{Exp}

We will use two data sets in this section as a running illustration, both obtained from the Stanford SNAP network database \cite{snapnets}. The first is a coauthorship network built from the DBLP computer science bibliography, and the second is a co-purchasing network from Amazon. The networks were originally analyzed by Yang and Leskovec \cite{Yang2012} in one of the first papers to systematically analyze the properties of ground-truth communities (abbreviated in figures as GTCs) in real-world networks. Both communities have ground-truth community labels: 13,477 ground-truth community labels in DBLP, defined as connected components of authors within the same publication venue; and 271,570 ground-truth community labels in Amazon, defined using product categories. Additionally, Yang and Leskovec labeled 5000 of the communities in each data set as ``best'' in terms of having community-like properties such as low conductance or high triangle-participation ratio. We computed the core-periphery decomposition for both networks using the iterative node dominance collapse algorithm described in Section \ref{Algorithm}. For the Amazon co-purchasing network, the periphery consisted of 70716 nodes (accounting for only 21\% of the nodes in the network), each of which were singletons, connected only to the core and not to other peripheral nodes. To allow further collapse, we re-computed the core using the 2-hop neighbor sets $\mathcal{N}_2[v]$ described in Section \ref{Equiv}. This yielded 193,195 nodes in the periphery (57.7\% of the nodes in the network), with 70716 peripheral components, of which 20136 were non-singletons (of varying sizes). All analysis presented below uses the regular node dominance collapse on the DBLP data set, and the node dominance collapse based on 2-hop neighbor sets for the Amazon data set.

Descriptive statistics for the networks, as well as for their associated core-periphery partitions, are presented in Table \ref{table:DescStats}. For the computations of average degree and clustering coefficient, the values were computed with respect to the entire network, and again with respect to the induced subgraph under consideration (either the core or periphery).

\begin{table}\label{table:DescStats}
\caption{Descriptive statistics for the DBLP and Amazon data sets, and their core-periphery decompositions.}
\begin{center}
\begin{tabular}{|lll|}
\hline
& DBLP & Amazon \\
\hline
Nodes in core: & 71,018 & 141,688 \\
Nodes in periphery: & 246,062 & 193,195 \\
Nodes (total): & \textbf{317,080} & \textbf{334,863} \\
\hline
Edges within core: & 318,741 & 347,527 \\
Edges within periphery: & 274,367 & 218,237 \\
Edges between core and periphery: & 456,758 & 360,108 \\
Edges (total): & \textbf{1,049,866} & \textbf{925,872} \\
\hline
Mean degree: & & \\
\hspace{2mm} Entire network & 6.62 & 5.53 \\
\hspace{2mm} Core (w.r.t entire network) & 15.41 & 7.45 \\
\hspace{2mm} Core (w.r.t. core) & 8.98 & 4.91 \\
\hspace{2mm} Periphery (w.r.t entire network) & 4.09 & 4.12 \\
\hspace{2mm} Periphery (w.r.t periphery) & 2.23 & 2.26 \\
\hline
Clustering coefficient: & & \\
\hspace{2mm} Entire network & 0.632 & 0.397 \\
\hspace{2mm} Core (w.r.t entire network) & 0.285 & 0.219 \\
\hspace{2mm} Core (w.r.t. core) & 0.255 & 0.182 \\
\hspace{2mm} Periphery (w.r.t entire network) & 0.733 & 0.527 \\
\hspace{2mm} Periphery (w.r.t periphery) & 0.385 & 0.293 \\
\hline
Communities (total): & & \\
\hspace{2mm} Number & 13,477 & 271,570 \\
\hspace{2mm} Average size & 53.41 & 11.67 \\
\hspace{2mm} Standard deviation of size & 257.58 & 273.66 \\
\hline
Communities (best): & & \\
\hspace{2mm} Number & 5000 & 5000  \\
\hspace{2mm} Average size & 22.45 & 13.49 \\
\hspace{2mm} Standard deviation of size & 201.08 & 17.52 \\
\hline
\end{tabular}
\end{center}
\end{table}

To verify the stability of the core under multiple realizations of the node dominance collapse algorithm, we performed the following randomization: For one realization of the iterated node dominance collapse, we would compute the set of dominated nodes, pick one at random to collapse, add the newly dominated nodes to the set of dominated nodes, randomly pick the next dominated node to collapse, and so on. After performing 100 realizations of the core-periphery decomposition on the two data sets, we found that 99.58\% (DBLP) and 99.43\% (Amazon) of the nodes in the core were present in the core on every realization. The set of nodes that appeared in the core on some (but not all) realizations was 0.89\% (DBLP) 1.24\% (Amazon) the size of the core. Thus, not only is the shape of the core unique, but the actual nodes composing it are very stable in these real-world data sets.

\subsection{Relationship of core-periphery to network structure}\label{ObsProp}

For both data sets, we observe (Table \ref{table:DescStats}) that nodes in the core have higher degree than nodes in the periphery, with the difference especially pronounced in the DBLP network. Additionally, nodes in the core have lower clustering coefficient, which corroborates our intuition that core nodes have ``diverse friend sets'', so their friends are not all friends with each other. Along with their high degree, this is also interpretable as having reach outside of their local community.

\begin{figure}[htp]
\begin{center}
\begin{tabular}{l}
\includegraphics[scale=0.25]{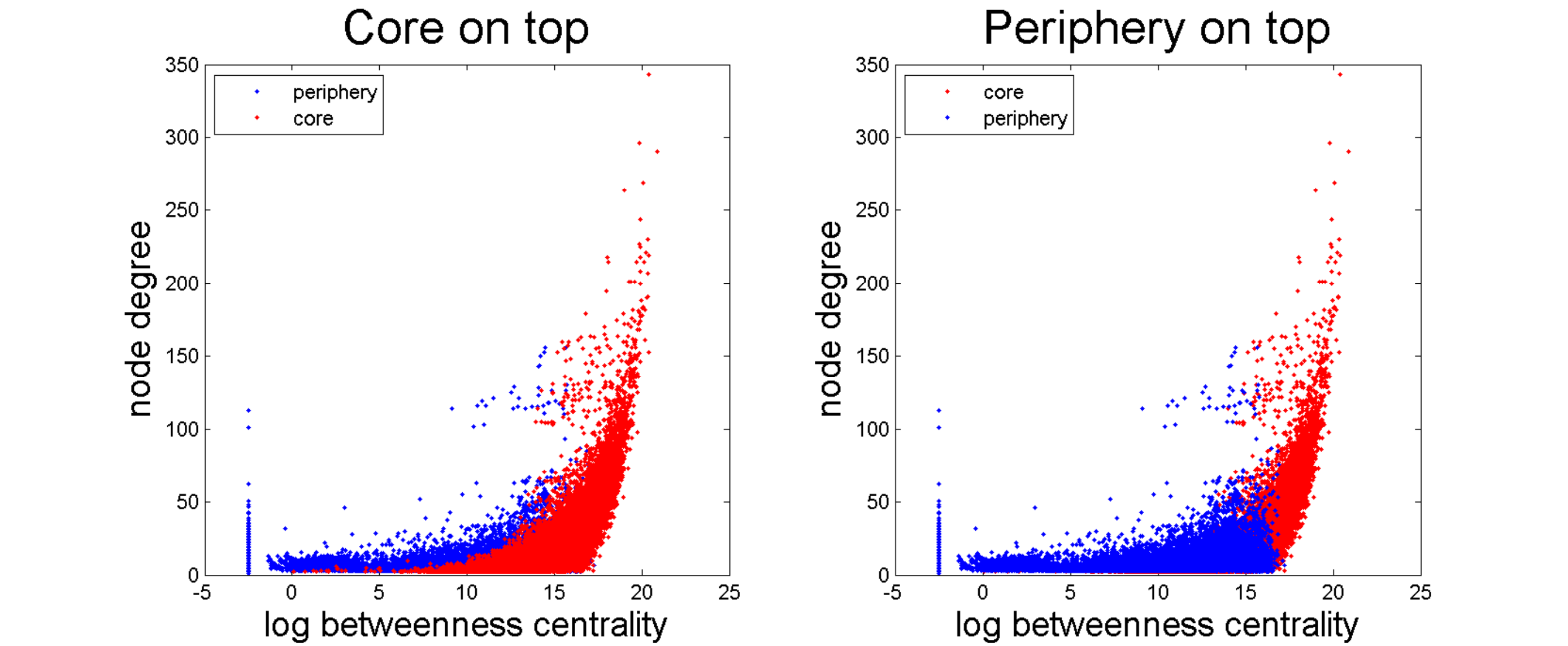} \\
\includegraphics[scale=0.275]{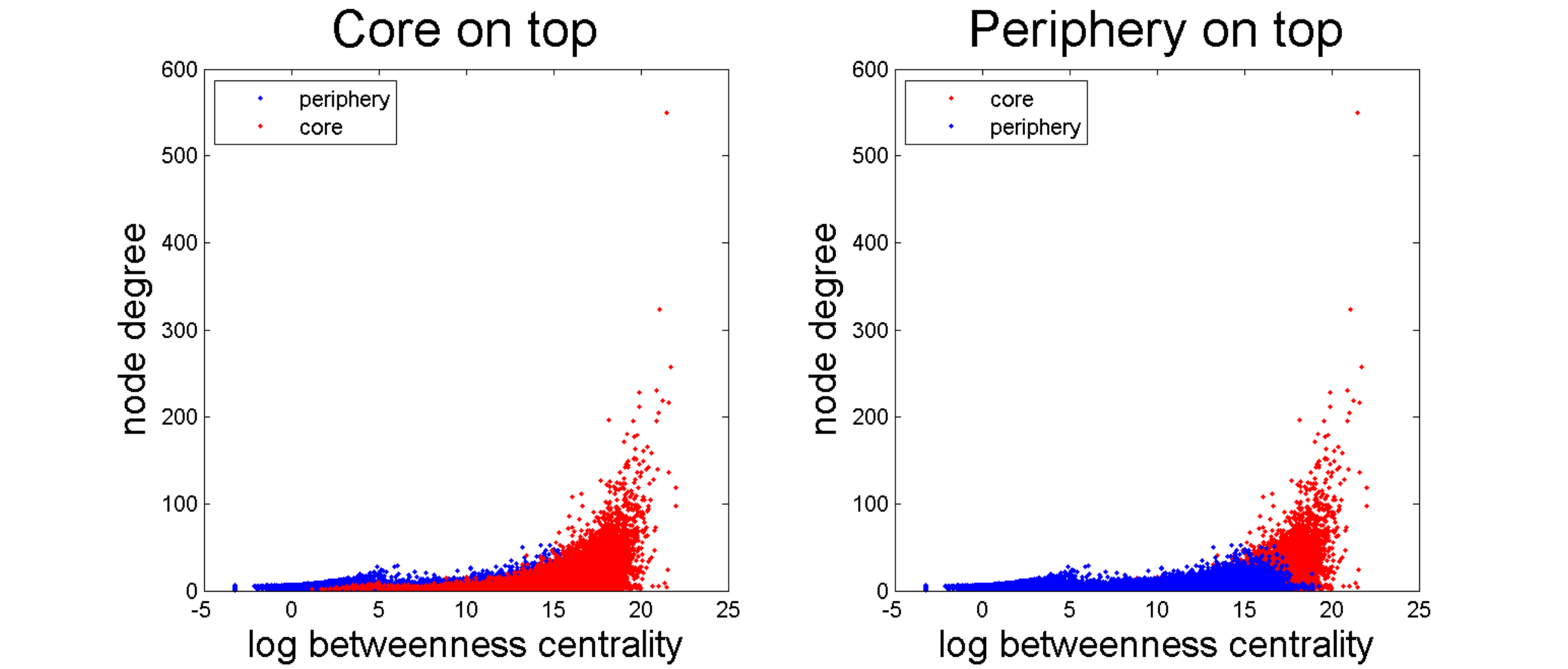} \\
\end{tabular}
\end{center}
\caption{Log betweenness centrality vs degree in core and periphery (DBLP-top, Amazon-bottom)\label{LogBCvsDegDBLP}}
\end{figure}

Scatterplots showing the natural logarithm of betweenness centrality versus node degree are shown in Figure \ref{LogBCvsDegDBLP}, with the two plots of the same data alternating whether core or periphery is plotted on top, to help display the region of overlap. As mentioned in Section \ref{Flow}, all nodes with betweenness centrality of zero (i.e. nodes through which no shortest paths pass) are guaranteed to be in the periphery, and we observe that additionally, all of the nodes with highest betweenness centrality are in the core. For example, in Figure \ref{LogBCvsDegDBLP}, it can be seen that in the DBLP data set there is a threshold betweenness centrality value (around $\ln($bc$)= 17$), above which all nodes are in the core, while in the Amazon data set, it is the nodes with both high degree and high betweenness centrality that appear exclusively in the core.

\begin{figure}[htp]
\begin{center}
\begin{tabular}{l}
\includegraphics[scale=0.25]{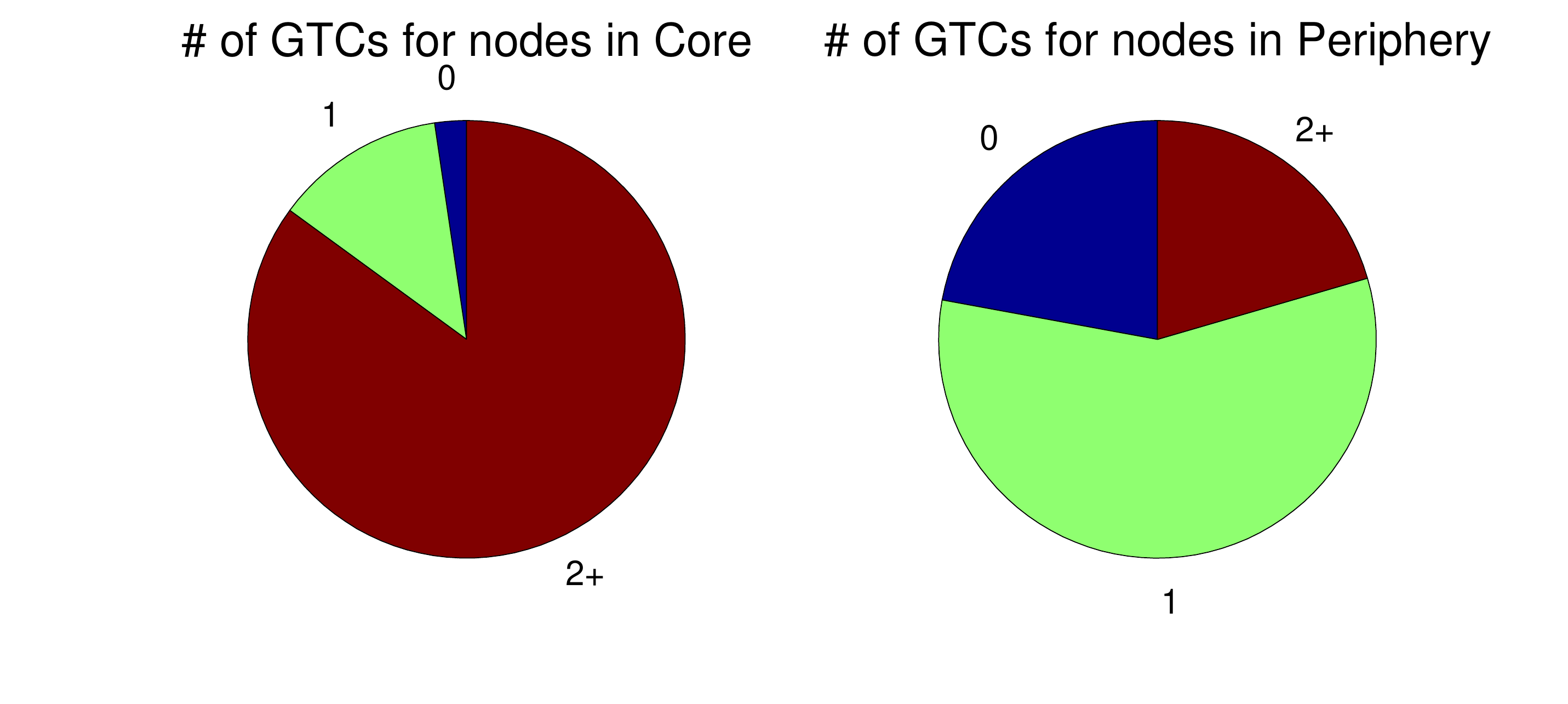} \\
\includegraphics[scale=0.25]{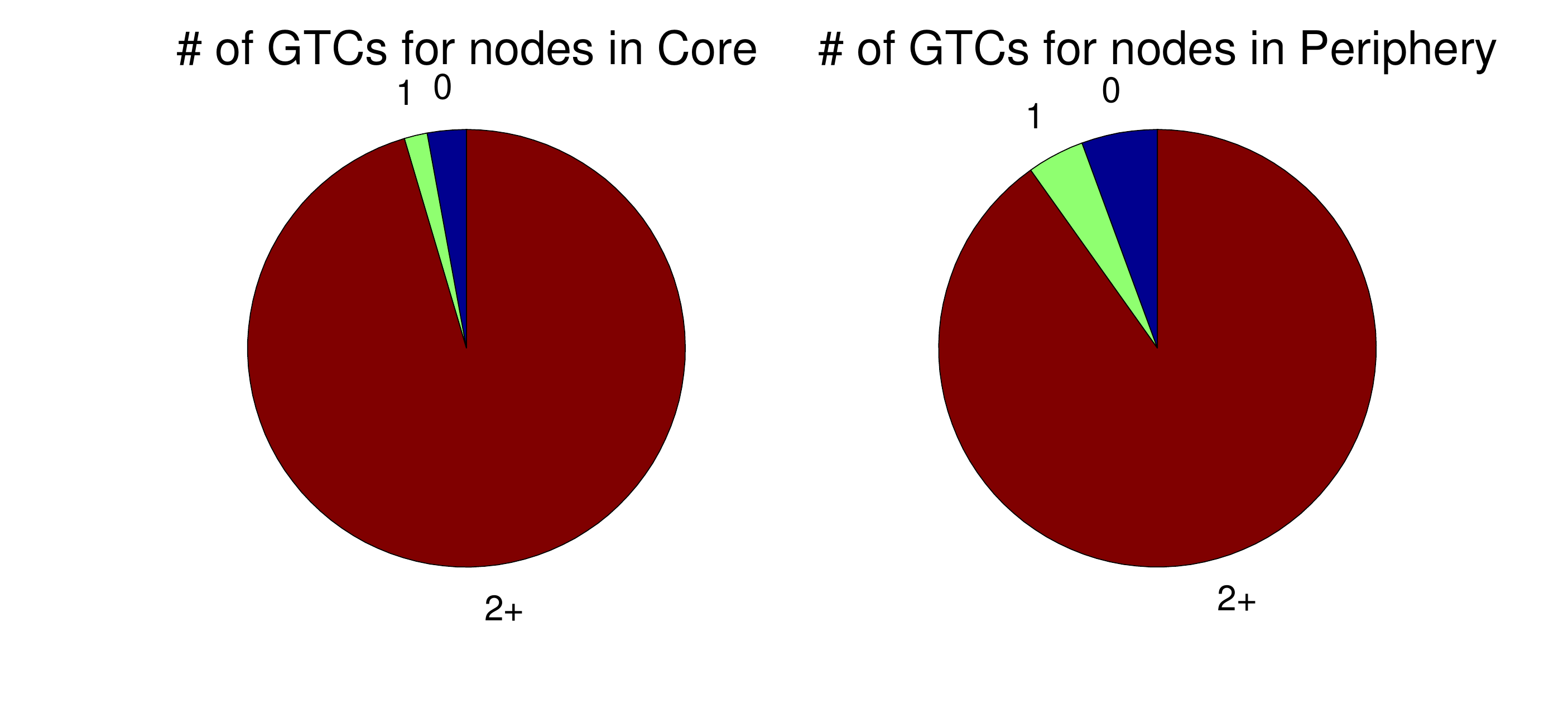} \\
\end{tabular}
\end{center}
\caption{Number of community memberships for nodes in core and periphery (DBLP-top, Amazon-bottom)\label{PieNumGTCsDBLP}}
\end{figure}

Figure \ref{PieNumGTCsDBLP} shows the number of ground-truth community assignments per node in the core and periphery of the DBLP and Amazon networks. Out of all the nodes in the periphery, 22.11\% had no ground-truth community (GTC) membership labels, 57.39\% had exactly one, and 20.49\% had more than one GTC membership label. On the other hand, out of the nodes in the core 85.02\% had multiple GTC membership labels, while  12.65\% had a single community, and only 2.33\% had no GTC label. From another perspective, the periphery contained 97.05\% of the nodes without a GTC label, 94.02\% of the nodes with a single label, but 45.51\% of the nodes with multiple labels (however of those nodes multiply labeled, the average number of labels was 2.9 in the periphery, but 7.0 in the core). A similar behavior is observed in the Amazon network, albeit to a lesser extent, and likely due to the average number of labels per node being much higher.

\subsection{Role of core in network flow}

To demonstrate the key role our core nodes play in information flow over the network, we computed their contribution to the shortest paths of the network. For each network, we randomly chose 1000 pairs of nodes, and computed shortest paths between them. Since 100\% of these paths contain at least one node from the core, we computed the proportion of each path that is in the core. For comparison, we chose three sets of nodes, each with the same number of nodes as the core: chosen uniformly randomly; using the nodes of highest degree; and using the nodes with highest betweenness centrality. Then, using the same 1000 shortest paths, we computed the proportion of nodes from each path belonging to each of these sets. Taking the average over all 1000 paths, the mean proportion of each path contained in the four sets (Core, Highest BC, Highest Degree, and Random) are shown in Table \ref{table:SPs}. Since betweenness centrality measures how many shortest paths pass through a node, the nodes with highest betweenness centrality should be the optimal choice for this measure (if considering all shortest paths in the entire network), so it is not surprising that they have the highest proportion of shortest path nodes. What is somewhat more surprising, is that for both data sets, the nodes in the core out-perform the nodes with highest degree, so a greater proportion of nodes in shortest paths belong to the core, than belong to the equal-sized set of highest degree nodes. The proportion of nodes in the shortest paths that belong to the Random set give us a baseline probability from which to compare the other choices of ``important'' nodes. Recall also, that betweennness centrality is very expensive computationally, requiring global information, so it is useful that the distributed core-periphery computation be nearly comparable at obtaining nodes central to network flow.

\begin{table}\label{table:SPs}
\caption{Importance of core nodes, high betweenness centrality nodes, high degree nodes, and randomly chosen nodes, in shortest paths of the DBLP and Amazon networks}
\begin{center}
\begin{tabular}{|l|ll|}
\hline
\multicolumn{3}{|c|}{Proportion of nodes in shortest paths} \\
\multicolumn{3}{|c|}{belonging to important sets} \\
\hline
& DBLP & Amazon \\
\hline
Highest BC & 0.785 & 0.892 \\
\textbf{Core} & \textbf{0.753} & \textbf{0.841} \\
Highest degree & 0.739 & 0.698 \\
Random & 0.222 & 0.427 \\
\hline
\end{tabular}
\end{center}
\end{table}

\subsection{Community detection}\label{CommDetec}

\begin{figure}[htp]
\begin{center}
\begin{tabular}{lll}
\includegraphics[scale=0.15]{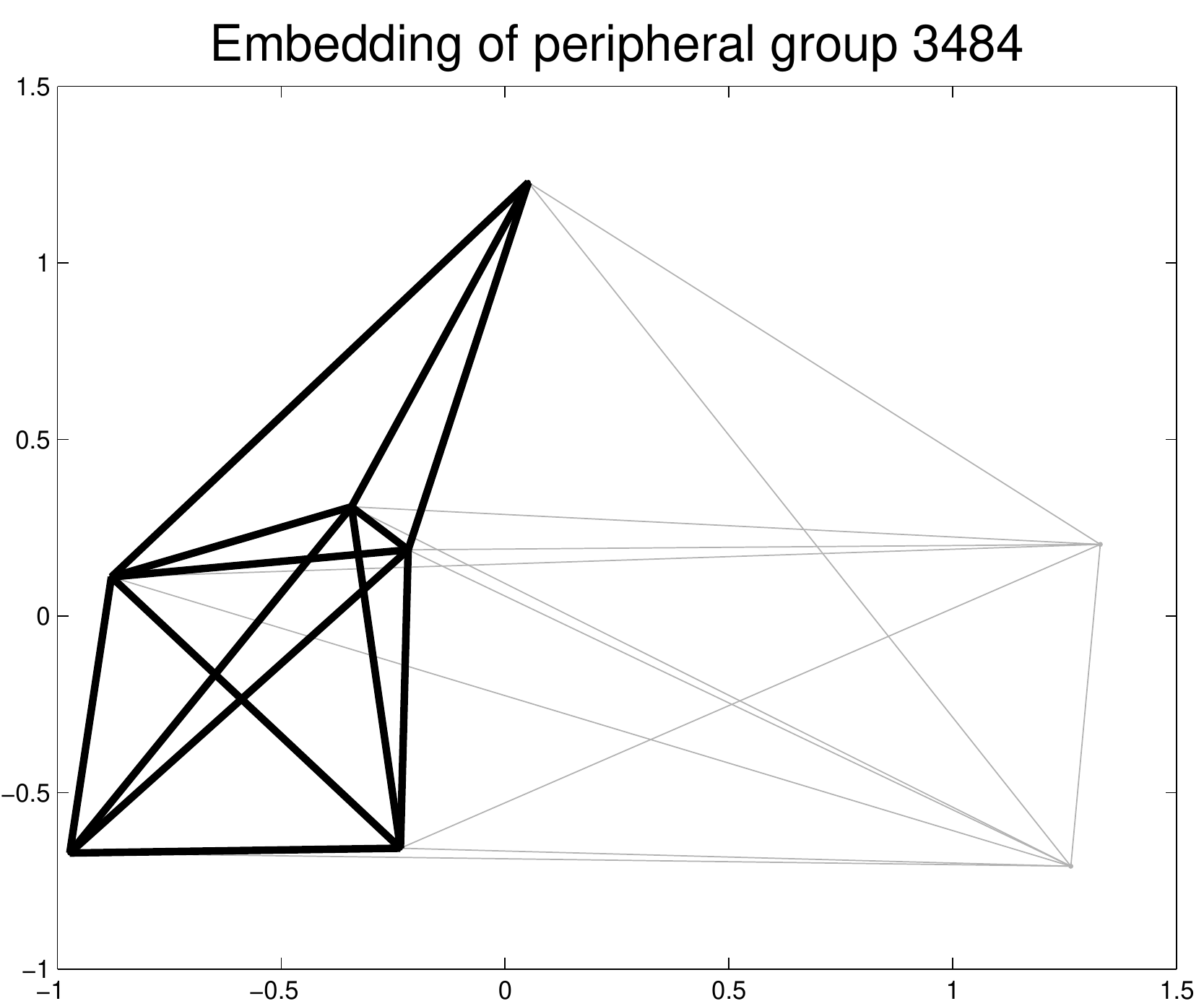} & \includegraphics[scale=0.15]{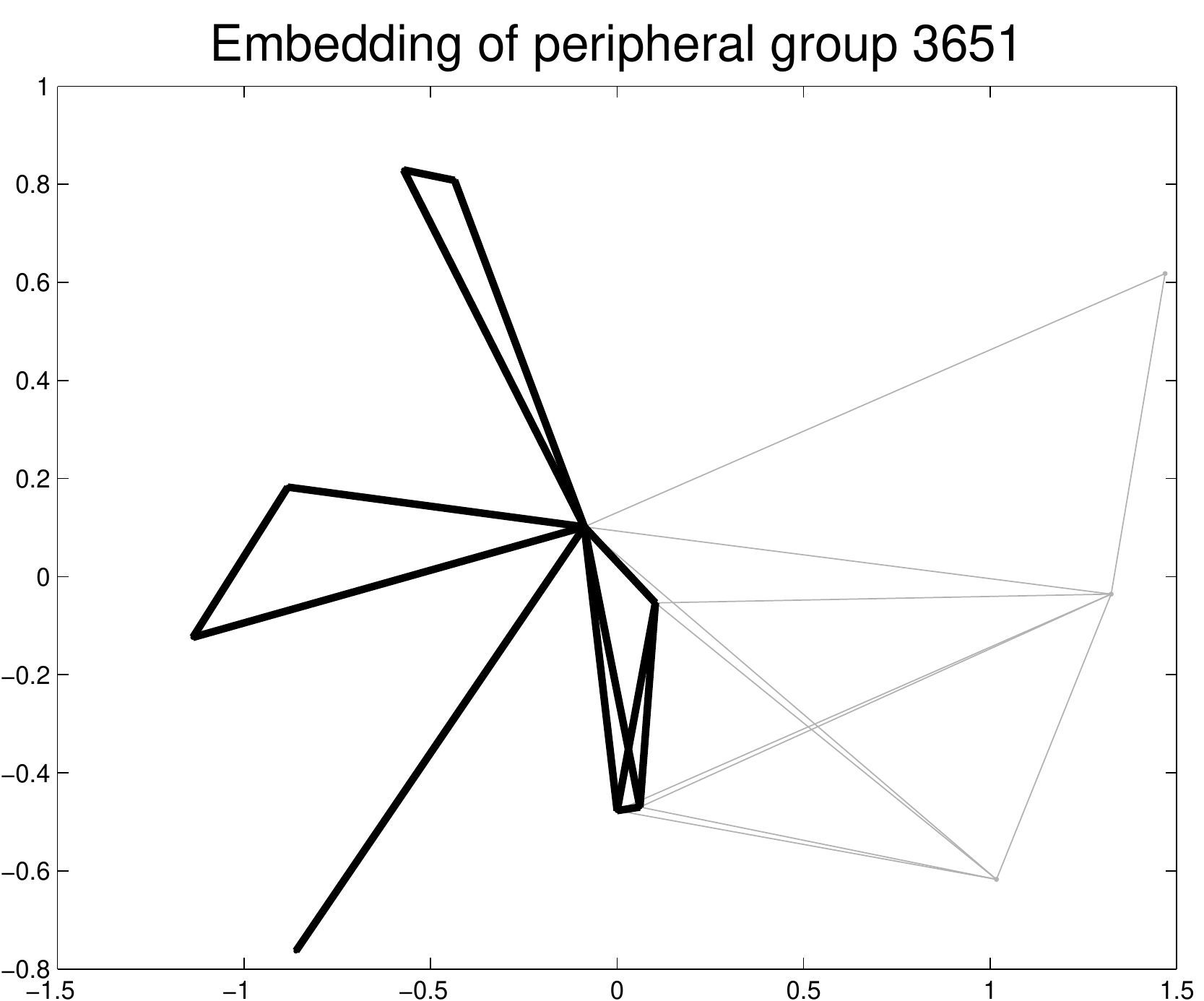} & \includegraphics[scale=0.15]{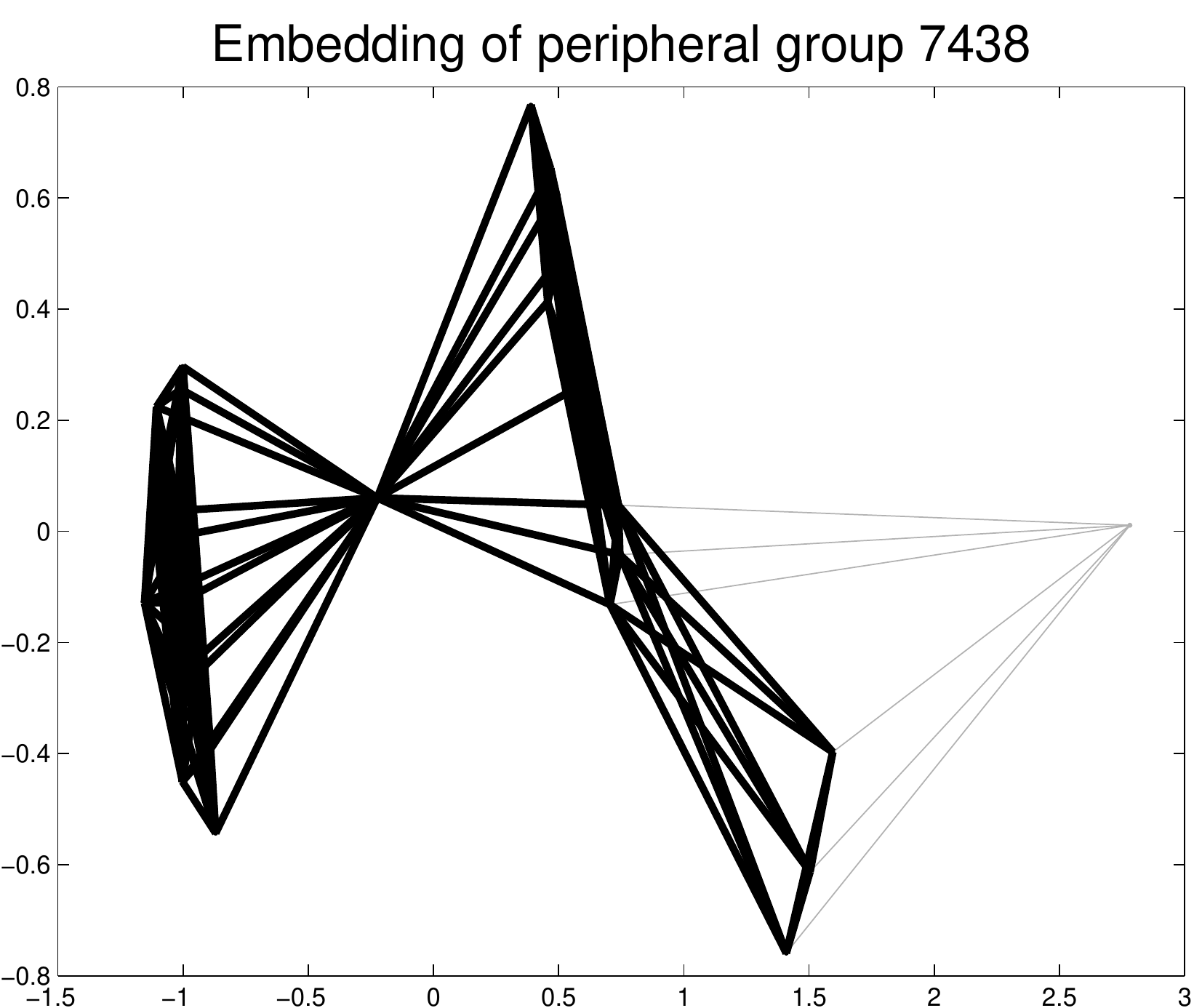}\\
\end{tabular}
\end{center}
\caption{Example peripheral components. \label{PCs}}
\end{figure}

The findings of this study are consistent with the community affiliation graph model (AGM) of Yang and Leskovec \cite{Yang2012b,Yang2014}, in the sense that it supports an overlapping community model for social and information networks where the probability of an edge between two nodes is related to their common community membership(s), with higher probabilities of edges between nodes that have multiple communities in common. Under this model, we showed that nodes are only dominated (with very high probability) by nodes which share their community memberships. Interpreting our peripheral components with respect to this model, they appear to be the `non-overlapping' parts of communities that stick out of the network. Figure \ref{PCs} shows embeddings of some peripheral components from the DBLP data set as examples, where the peripheral component is drawn in black, while the core nodes and connecting edges are grey. The internal structure and connectivity to the core can vary considerably between peripheral components.

In light of the interpretation of peripheral components as non-overlapping portions of communities, we propose an algorithm which consists of taking unions of these peripheral components, along with their neighboring nodes in the core, to obtain candidate sets for community detection.

More precisely, let $PC = \{pc_i\}_{i=1}^{|PC|}$ denote the set of peripheral components in the network, where each node in the periphery is in exactly one peripheral component, $pc_i$. Then define the extended peripheral components $PC^+ = \{pc_i^+\}_{i=1}^{|PC|}$ where
\[ pc_i^+ = \{ v \in V_c \mbox{ } | \mbox{ } \exists \mbox{ } v_j \in pc_i \mbox{ s.t. } (v_j,v) \in E \} \cup pc_i, \]
so each extended peripheral component additionally contains all the nodes in the core that share an edge with a vertex of the peripheral component. The extended peripheral components are meant to approximate ground-truth communities in the data set, however there are large numbers of very small size (such as those consisting of an isolated peripheral node and its single neighboring core node). We consolidate extended peripheral components into ``candidate sets'' by taking, for each $v \in V_C$, the union of all extended peripheral groups that include $v$. So we obtain $\{cs_v\}_{v \in V_C}$, where
\[ cs_v = \bigcup_{\substack{pc_i^+ \in PC^+ \\ v \in pc_i^+}} pc_i^+. \]
For example, if there were many peripheral nodes connected to a single core node (but not connected amongst each other), this group would be consolidated into a single candidate set. We then remove any candidate sets $cs_v$ that are repetitions or subsets of other candidate sets, to obtain our final set of maximal candidate sets: $CS$. 
Intuitively, our candidate sets are meant to approximate ground truth communities, or unions of ground truth communities (that overlap on common core nodes).

To judge the performance of our candidate sets for the purposes of community detection, we also ran the BIGCLAM algorithm \cite{Yang2013} on the DBLP data set. Popular methods for detection overlapping communities include clique percolation, link clustering, and fuzzy detection methods using mixed-membership stochastic block models (see \cite{Xie2013} for a survey), however none of these methods scale up well to networks with hundreds of thousands or millions of nodes. The recent exception to this is Yang and Leskovec's BIGCLAM algorithm, which can estimate the overlapping community structure for large networks. The BIGCLAM algorithm (available in the SNAP C++ package \cite{snap}) allows the user to input the expected number of communities, but runs into memory problems if the number of communities is larger than a few hundred. It also has an option for the algorithm to learn the appropriate number of communities, with a default to test between 5 and 100 communities. Therefore, to obtain a set of communities of the same order as the number of ground-truth communities (13,477 for the DBLP data set), we performed BIGCLAM in a nested manner. First obtaining 100 communities, and then further subdividing each of these, where the optimal number of subcommunities was most often also 100. This yielded a total of 9904 detected communities from the BIGCLAM algorithm. We used the same method for analysis of the Amazon data set, yielding 8899 BIGCLAM communities, even though that network has a much larger number of ground-truth communities (271,570). For both data sets, the number of candidate sets obtained using our method was around 40,000 (47,134 for DBLP and 37,449 for Amazon).

To measure the fit of the candidate sets and BIGCLAM communities to the ground-truth communities, we used precision, recall, and average F1 score. For a detected community $C_1$ and ground truth community $C_2$ (the target), the \emph{precision} is the proportion of detected nodes that belong to the target:
\[ precision(C_1,C_2) = \frac{|C_1 \cap C_2|}{|C_1|}, \]
the \emph{recall} is the proportion of target nodes captured in the detected community:
\[ recall(C_1,C_2) = \frac{|C_1 \cap C_2|}{|C_2|}, \]
and the F1-score is the harmonic mean of precision and recall:
\[ F1(C_1,C_2) = \frac{precision(C_1,C_2) \cdot recall(C_1,C_2)}{2(precision(C_1,C_2)+recall(C_1,C_2))}. \]
These three values for a given ground-truth community are obtained by maximizing each over all candidate sets (BIGCLAM communities), and an average precision, recall, and F1-score for the ground-truth communities is obtained. Similarly, the three values are obtained for each candidate set (BIGCLAM community) by thinking of it as the ``target'' community, and maximizing precision, recall, and F1-score over all ground-truth communities, and then taking the average of these maxima.

Using all three of these values (precision, recall, and F1-score) helps offset some of the discrepancies caused by the varying numbers of ground-truth communities, candidate sets, and BIGCLAM communities. Since the matching of ground-truth communities onto detected communities, but also the matching of detected communities onto ground-truth communities, are considered, having more candidate sets than BIGCLAM communities will not necessarily be an advantage.

Table \ref{table:detectionAll} gives the values for recall, precision and F1-score when comparing the ground-truth communities to our candidate sets (left three columns), and to the BIGCLAM communities (right three columns). The performance using candidate sets and BIGCLAM communities are compared for each measure (eg. ``ground-truth community recall'', or `` average precision''), with the values in boldface indicating the method (candidate sets or BIGCLAM) with superior performance in that measure. The column ``ground-truth'' gives the average values for the ground truth communities (when maximized over the detected communities), and the column ``detected'' gives the average for the detected communities (when maximized over ground-truth communities).

Our candidate sets give better overall community detection performance than the BIGCLAM communities (as measured by the average F1-score). For the DBLP data set, the ground-truth communities were contained in the candidate sets (based on higher ground-truth recall scores), more so than the candidate sets found strongly-matching ground-truth communities (although it is worth noting, as Yang and Leskovec did, that not all ``true'' ground-truth communities necessarily have ground-truth community labels in this data set). The performance on the Amazon data set is quite good, with very high ground-truth recall and detected recall and precision for both the candidate sets and the BIGCLAM methods, although our candidate sets out-performed BIGCLAM in detected recall, as well as ground-truth, detected and average F1-scores.

\begin{table}\label{table:detectionAll}
\renewcommand{\tabcolsep}{2pt}
\caption{Detection of all ground-truth communities by candidate sets and BIGCLAM communities}
\begin{center}
\begin{tabular}{|l|ccc|ccc|}
\hline
& \multicolumn{6}{|c|}{DBLP (all 13,477 communities)} \\
\hline
& \multicolumn{3}{c|}{Candidate sets} & \multicolumn{3}{|c|}{BIGCLAM} \\
\hline
& ground-truth & detected & average & ground-truth & detected & average \\
\hline
Recall & \textbf{0.7620} & \textbf{0.5401} & \textbf{0.6511} & 0.7418 & 0.4478 & 0.5948 \\
Precision & \textbf{0.4319} & 0.4960 & \textbf{0.4640} & 0.2366 & \textbf{0.6261} & 0.4314 \\
F1-score & \textbf{0.4233} & 0.2565 & \textbf{0.3399} & 0.2696 & \textbf{0.2721} & 0.2709 \\
\hline
\multicolumn{7}{c}{} \\
\hline
& \multicolumn{6}{|c|}{Amazon (all 271,570 communities)} \\
\hline
& \multicolumn{3}{c|}{Candidate sets} & \multicolumn{3}{|c|}{BIGCLAM} \\
\hline
& ground-truth & detected & average & ground-truth & detected & average \\
\hline
Recall & 0.8481 & \textbf{0.8721} & 0.8601 & \textbf{0.9213} & 0.8203 & \textbf{0.8708} \\
Precision & \textbf{0.2545} & 0.8728 & \textbf{0.5636} & 0.1124 & \textbf{0.9861} & 0.5492 \\
F1-score & \textbf{0.3218} & \textbf{0.4815} & \textbf{0.4017} & 0.1611 & 0.4685 & 0.3148 \\
\hline
\multicolumn{7}{c}{} \\
\hline
& \multicolumn{6}{|c|}{DBLP (5000 best communities)} \\
\hline
& \multicolumn{3}{c|}{Candidate sets} & \multicolumn{3}{|c|}{BIGCLAM} \\
\hline
& ground-truth & detected & average & ground-truth & detected & average \\
\hline
Recall & \textbf{0.9414} & 0.2559 & \textbf{0.5987} & 0.9054 & \textbf{0.2678} & 0.5866 \\
Precision & \textbf{0.4313} & 0.3121 & \textbf{0.3717} & 0.3065 & \textbf{0.4216} & 0.3640 \\
F1-score & \textbf{0.5221} & 0.1446 & \textbf{0.3333} & 0.3840 & \textbf{0.1913} & 0.2877 \\
\hline
\multicolumn{7}{c}{} \\
\hline
& \multicolumn{6}{|c|}{Amazon (5000 best communities)} \\
\hline
& \multicolumn{3}{c|}{Candidate sets} & \multicolumn{3}{|c|}{BIGCLAM} \\
\hline
& ground-truth & detected & average & ground-truth & detected & average \\
\hline
Recall & \textbf{0.9893} & 0.0222 & \textbf{0.5058} & 0.9072 & \textbf{0.0728} & 0.4900 \\
Precision & \textbf{0.4781} & 0.0404 & 0.2593 & 0.4535 & \textbf{0.1224} & \textbf{0.2880} \\
F1-score & \textbf{0.5753} & 0.0241 & \textbf{0.2997} & 0.5100 & \textbf{0.0753} & 0.2927 \\
\hline
\end{tabular}
\end{center}
\end{table}

The analysis was repeated using only the 5000 ``best'' ground-truth communities, and again the candidate sets resulted in higher average F1-scores than the BIGCLAM communities. The main difference was that recall for the ground-truth communities increased (on average, each ground-truth community had a candidate set it was 94\% contained in), while recall and precision for the candidate sets decreased (since there were fewer ground-truth communities to match to, fewer detected had a well-matched ground-truth community). It is also worth noting that for the DBLP data set 81.7\% of the best ground-truth communities were completely contained in at least one candidate set, while 73.8\% of the best ground-truth communities were completely contained in at least one BIGCLAM community. For the Amazon data set, these values were 94.8\% for the candidate sets, and 82.8\% for the BIGCLAM communities.

The challenge of detecting thousands of overlapping communities from a large network is formidable. Currently there are no available methods which achieve excellent performance when comparing detected to ground-truth communities. Based on the analysis of two large, real-world data sets with ground-truth community information, our proposed algorithm of obtaining candidate sets from the peripheral components of the core-periphery decomposition, yielded more accurate community detection results than the state-of-the-art BIGCLAM algorithm for overlapping community detection, with much lower complexity and a distributed algorithm.

\section{Conclusion}\label{CorePeriphConcl}
This study posed the question ``How does the concept of node dominance relate to local and global properties of a network?''. Previous work determined that iteratively removing dominated nodes is a homology-preserving way to perform a collapse/simplification of a simplicial complex \cite{Barmak2012} \cite{Wilkerson2013}. This was extended into a distributed algorithm for the case of flag complexes \cite{Wilkerson2013b}. Here, we undertook an investigation of the theoretical and practical properties of performing such a collapse on social and information networks, and discovered that it has implications for both a core-periphery decomposition of the network, as well as uncovering network community structure.

The properties of the core and periphery that we developed in Section \ref{Properties}, and observed in Section \ref{Exp}, lead to the interpretation that nodes in the core obtained using node dominance collapse are important with respect to network flow, to the global structure of the network, and to the network community structure.

The core nodes are essential to network flow because of two properties: a shortest path between any two points in the core is contained in the core; and nodes with betweenness centrality zero (through which no shortest paths pass) are never in the core. Observationally, `hub' nodes are contained in the core, and core nodes often have high degree and high betweenness centrality.

The global structure of the network is preserved in the core because the homology of the core is the same as the homology of the entire network, when considering the respective flag complexes. This can be interpreted as node dominance collapses only having `local' effects, and that nodes with diverse neighbor sets (including bridging ties) are members of the core, maintaining a scaffolding for the global structure of the network. The observation that each core node typically has a diverse neighbor set (their friends are not all friends with each other) is also quantified by their relatively low clustering coefficient values.

Finally, the core is related to the community structure of the network because under community membership models where within-community connections have significantly higher probability than cross-community connections, we see that nodes are dominated (with high probability) by nodes that share their community membership(s). In real-world networks with overlapping ground-truth community labels, this is observed through nodes with multiple community memberships typically residing in the core, and through nodes with single (or no) community labels occupying the periphery.

The result relating the core-periphery to the community structure of the network gives us an additional application: the use of the peripheral components to generate ``candidate sets'' which are likely to contain the true network communities. Many state-of-the-art community detection algorithms which allow for overlapping communities, are not scalable past network sizes of a few thousand nodes. The notable recent exception is Yang and Leskovec's BIGCLAM algorithm, which our method is shown to outperform on their DBLP dataset.

Implications of this work may be of interest not only to researchers explicitly interested in a core-periphery decomposition of complex networks, but to anyone studying community structure, or key nodes for network flow. Hopefully this work will also serve to further popularize the node dominance collapse for use in general contexts where data is represented using a simplicial complex structure.

One limitation of our method is that some networks don't collapse using node dominance. For example, on Facebook there are very few people who have a friend list completely contained in the friend list of another person. One option for future research in this direction would involve performing the node dominance collapse locally on ego networks, and consolidating the resulting communities. Another potential drawback is the nondeterministic nature of the node dominance collapse algorithm. Perhaps under some circumstances it would be wise to consider the set of nodes that are ``ever in the core'', or ``always in the core'', under repeated realizations of the algorithm. In practice however (Section \ref{ObsProp}), we have seen that these two sets are quite similar.

One other area for future research is in the study of the core under a graph evolution. Either using observed or model-generated dynamic networks, studying how the core varies over time could be used to help evaluate or predict community structure and key players in the network.


%

%

%
%

\ifCLASSOPTIONcaptionsoff
  \newpage
\fi



\bibliographystyle{IEEEtran}
\bibliography{TSIPNBib}
%
%
%

%

\begin{IEEEbiography}{Jennifer Gamble}
Biography text here.
\end{IEEEbiography}

\begin{IEEEbiography}{Harish Chintakunta}
Biography text here.
\end{IEEEbiography}

\begin{IEEEbiography}{Adam Wilkerson}
Biography text here.
\end{IEEEbiography}

\begin{IEEEbiography}{Terrence J. Moore}
Biography text here.
\end{IEEEbiography}

\begin{IEEEbiography}{Ananthram Swami}
Biography text here.
\end{IEEEbiography}

\begin{IEEEbiography}{Hamid Krim}
Biography text here.
\end{IEEEbiography}






\end{document}